\documentclass[a4paper]{jpconf}
\usepackage{graphicx}

\begin{document}
\title{Multiplicity fluctuations \\ in nuclear collisions at ${\bf 158\,{\it A}}$~GeV}

\author{M Rybczy\'nski$^{12}$, C~Alt$^{9}$, T~Anticic$^{21}$,
B~Baatar$^{8}$,D~Barna$^{4}$, J~Bartke$^{6}$, L~Betev$^{9,10}$,
H~Bia{\l}\-kowska$^{19}$, A~Billmeier$^{9}$, C~Blume$^{9}$,
B~Boimska$^{19}$, M~Botje$^{1}$, J~Bracinik$^{3}$, R~Bramm$^{9}$,
R~Brun$^{10}$, P~Bun\v{c}i\'{c}$^{9,10}$, V~Cerny$^{3}$,
P~Christakoglou$^{2}$, O~Chvala$^{15}$, J~G~Cramer$^{17}$,
P~Csat\'{o}$^{4}$, N~Darmenov$^{18}$, A~Dimitrov$^{18}$,
P~Dinkelaker$^{9}$, V~Eckardt$^{14}$, G~Farantatos$^{2}$,
D~Flierl$^{9}$, Z~Fodor$^{4}$, P~Foka$^{7}$, P~Freund$^{14}$,
V~Friese$^{7}$, J~G\'{a}l$^{4}$, M~Ga\'zdzicki$^{9,12}$,
G~Georgopoulos$^{2}$, E~G{\l}adysz$^{6}$, K~Grebieszkow$^{20}$,
S~Hegyi$^{4}$, C~H\"{o}hne$^{13}$, K~Kadija$^{21}$,
A~Karev$^{14}$, M~Kliemant$^{9}$, S~Kniege$^{9}$,
V~I~Kolesnikov$^{8}$, T~Kollegger$^{9}$, E~Kornas$^{6}$,
R~Korus$^{12}$, M~Kowalski$^{6}$, I~Kraus$^{7}$, M~Kreps$^{3}$,
M~van~Leeuwen$^{1}$, P~L\'{e}vai$^{4}$, L~Litov$^{18}$,
B~Lungwitz$^{9}$, M~Makariev$^{18}$, A~I~Malakhov$^{8}$,
C~Markert$^{7}$, M~Mateev$^{18}$, B~W~Mayes$^{11}$,
G~L~Melkumov$^{8}$, C~Meurer$^{9}$, A~Mischke$^{7}$,
M~Mitrovski$^{9}$, J~Moln\'{a}r$^{4}$, S~Mr\'owczy\'nski$^{12}$,
G~P\'{a}lla$^{4}$, A~D~Panagiotou$^{2}$, D~Panayotov$^{18}$,
A~Petridis$^{2}$, M~Pikna$^{3}$, L~Pinsky$^{11}$,
F~P\"{u}hlhofer$^{13}$, J~G~Reid$^{17}$, R~Renfordt$^{9}$,
A~Richard$^{9}$, C~Roland$^{5}$, G~Roland$^{5}$,
A~Rybicki$^{6,10}$, A~Sandoval$^{7}$, H~Sann$^{7}$,
N~Schmitz$^{14}$, P~Seyboth$^{14}$, F~Sikl\'{e}r$^{4}$,
B~Sitar$^{3}$, E~Skrzypczak$^{20}$, G~Stefanek$^{12}$,
R~Stock$^{9}$, H~Str\"{o}bele$^{9}$, T~Susa$^{21}$,
I~Szentp\'{e}tery$^{4}$, J~Sziklai$^{4}$, T~A~Trainor$^{17}$,
V~Trubnikov$^{20}$, D~Varga$^{4}$, M~Vassiliou$^{2}$,
G~I~Veres$^{4,5}$, G~Vesztergombi$^{4}$, D~Vrani\'{c}$^{7}$,
A~Wetzler$^{9}$, Z~W{\l}odarczyk$^{12}$ I~K~Yoo$^{16}$,
J~Zaranek$^{9}$, J~Zim\'{a}nyi$^{4}$}
\begin{center}
(NA49 Collaboration)
\end{center}
\address{$^{1}$NIKHEF, Amsterdam, Netherlands. \\
$^{2}$Department of Physics, University of Athens, Athens, Greece.\\
$^{3}$Comenius University, Bratislava, Slovakia.\\
$^{4}$KFKI Research Institute for Particle and Nuclear Physics, Budapest, Hungary.\\
$^{5}$MIT, Cambridge, USA.\\
$^{6}$Institute of Nuclear Physics, Cracow, Poland.\\
$^{7}$Gesellschaft f\"{u}r Schwerionenforschung (GSI), Darmstadt, Germany.\\
$^{8}$Joint Institute for Nuclear Research, Dubna, Russia.\\
$^{9}$Fachbereich Physik der Universit\"{a}t, Frankfurt, Germany.\\
$^{10}$CERN, Geneva, Switzerland.\\
$^{11}$University of Houston, Houston, TX, USA.\\
$^{12}$Institute of Physics \'Swi{\,e}tokrzyska Academy, Kielce, Poland.\\
$^{13}$Fachbereich Physik der Universit\"{a}t, Marburg, Germany.\\
$^{14}$Max-Planck-Institut f\"{u}r Physik, Munich, Germany.\\
$^{15}$Institute of Particle and Nuclear Physics, Charles University, Prague, Czech Republic.\\
$^{16}$Department of Physics, Pusan National University, Pusan, Republic of Korea.\\
$^{17}$Nuclear Physics Laboratory, University of Washington, Seattle, WA, USA.\\
$^{18}$Atomic Physics Department, Sofia University St. Kliment Ohridski, Sofia, Bulgaria.\\
$^{19}$Institute for Nuclear Studies, Warsaw, Poland.\\
$^{20}$Institute for Experimental Physics, University of Warsaw, Warsaw, Poland.\\
$^{21}$Rudjer Boskovic Institute, Zagreb, Croatia.}

%\address{Institute of Physics, \'Swi\c etokrzyska Academy,
%ul. \'Swi\c etokrzyska 15, PL - 25-406 Kielce, Poland}

\ead{mryb@pu.kielce.pl}

%%%%%%%%%%%%%%%

\begin{abstract}
System size dependence of multiplicity fluctuations of charged
particles produced in nuclear collisions at $158\,A$~GeV was
studied in the NA49 CERN experiment. Results indicate a
non-monotonic dependence of the scaled variance of the
multiplicity distribution with a maximum for semi-peripheral Pb+Pb
interactions with number of projectile participants of about 35.
This effect is not observed in a string-hadronic model of nuclear
collision HIJING.
\end{abstract}.

%%%%%%%%%%%%%%%

\section{Introduction}

Nucleus-nucleus collisions at relativistic energies have been
intensely studied over the last two decades. The main goal of
these efforts is to understand the properties of strongly
interacting matter under extreme conditions of high energy and
baryon densities when the creation of the quark-gluon plasma (QGP)
is expected \cite{Collins,Shuryak}. In fact, various collision
characteristics and their collision energy dependence suggest that
a transient state of deconfinement matter may be created at
collision energies as low as 30 A GeV \cite{Gazdzicki}.
Fluctuations in physical observables in heavy ion collisions have
been a topic of interest for some years as they may provide
important signals regarding the formation of QGP. With the large
number of particles produced in heavy ion collisions at CERN SPS
and BNL RHIC energies it has now become feasible to study
fluctuations on an event-by-event basis \cite{Heiselberg}. In a
thermodynamical picture of the strongly interacting system formed
in the collision, the fluctuations in particle multiplicities
\cite{WA98,Afanasev:2000fu,Roland:2004pu}, mean transverse momenta
\cite{KP}, and other global observables, are related to the
fundamental properties of the system, such the specific heat
\cite{Stodolsky:1995ds,Stephanov:1999zu}, chemical potential, and
matter compressibility \cite{Mrowczynski:1997kz}. These, in turn,
may reveal information on the properties of the equation of state
near the QCD phase boundary
\cite{Stephanov:1998dy,Stephanov:1999zu,Gazdzicki:2003bb}.

The main objective of this work is to study how the multiplicity
fluctuations change with increasing number of nucleons
participating in the collision, i.e. with the system size. In view
of this, centrality selected Pb+Pb collisions, central C+C and
Si+Si collisions as well as inelastic p+p interactions at
$158\,A$~GeV registered by NA49 at the CERN SPS were analyzed and
the results are presented.

The paper is organized as follows. In Sec.~\ref{s:method} the
method of measuring multiplicity fluctuations is introduced and
shortly discussed. The NA49 set-up is presented in
Sec.~\ref{s:experiment}. Experimental procedures, in particular
event and particle selection, detector acceptance and centrality
determination are discussed in Sec.~\ref{s:data}. The results on
the system size dependence of the multiplicity fluctuations are
presented in Sec.~\ref{s:results}. A final discussion and summary
close the paper.

%%%%%%%%%%%%%%%%%%%%%%

\section{Multiplicity fluctuations}
\label{s:method}
%%%%%%%%%%%%%%%%%%%%%%%%%

\subsection{Observables}

Let $P\left(N\right)$ be the multiplicity distribution, then:
\begin{equation}
\langle N\rangle= \sum N \cdot P(N)
\end{equation}
is the mean value of the distribution. The variance of the
multiplicity distribution is defined as:
\begin{equation}
Var\left(N\right)\equiv\sum\left(N-\langle
N\rangle\right)^{2}P\left(N\right)=\langle N^{2}\rangle -\langle
N\rangle^{2}.
\end{equation}
Note, that for the Poisonnian distribution the variance equals
mean value, $Var\left(N\right)=\langle N\rangle$.

\noindent Mean value and variance of the multiplicity
distributions are the only observables used for this analysis.

%%%%%%%%%%%%%%%%%%%%%%%%%%%

\subsection{Participants and spectators}

In description of nuclear collision the concepts of participant
and spectator nucleons are very useful. Participant nucleons are
nucleons which are removed by the interaction process from the
Fermi spheres of the target and projectile nuclei. The remaining
nucleons are called spectators. In case of central nucleus-nucleus
collisions, where the impact parameter $b$ is relatively small,
almost all nucleons participate in the collision. In particular,
the number of projectile participants, $N_{P}^{PROJ}$
approximately equals the total number of projectile nucleons:
$N_{P}^{PROJ}\simeq A$. If the collision is peripheral (with large
impact parameter $b$) almost all nucleons are spectators,
$N_{P}^{PROJ}<<A$. In general, the number of projectile spectators
$N_{SPEC}^{PROJ}$ is given by: $N_{SPEC}^{PROJ}=A - N_{P}^{PROJ}$.

%%%%%%%%%%%%%%%%%%%%%%%%%%%%%%%%%%%

\subsection{Multiplicity fluctuations in superposition model}

Widely used models of nuclear collisions are superposition models
which assume that secondary particles are emitted by independent
sources. The simplest example is the wounded nucleon model
\cite{Bialas:1976}, in which the sources are wounded nucleons,
i.e. the nucleons that have interacted within a Glauber model
approach. In this model the number of participants is equal to the
number of wounded nucleons. In superposition models the total
multiplicity is then given by:
\begin{equation}
N=\sum_{i=1}^{N_{S}}m_{i},
\end{equation}
where $N_{S}$ denotes number of sources and $m_{i}$ describes the
multiplicity from a single source. The mean total multiplicity in
an event may be calculated as:
\begin{equation}
\langle N\rangle=\langle N_{S}\rangle \cdot \langle m\rangle,
\label{nsr}
\end{equation}
where $\langle N_{S}\rangle$ is the mean number of sources and
$\langle m\rangle$ is the mean multiplicity from a single source.
The variance of the multiplicity distribution is given by:
\begin{equation}
Var(N)=\langle N_{S}\rangle \cdot Var(m)+\langle m\rangle^{2}\cdot
Var(N_{S}),
\label{var}
\end{equation}
where $Var(m)$ and $Var(N_{S})$ denote the variances of the
distribution of multiplicity from a single source and the
distribution of the number of sources, respectively.

\noindent The scaled variance of the multiplicity distribution
$Var(N)/\langle N\rangle$ is a useful measure of multiplicity
fluctuations. From Eqs.~\ref{nsr}~and~\ref{var} one gets
\begin{equation}
\frac{Var(N)}{\langle N\rangle}=\frac{Var(m)}{\langle
m\rangle}+\langle m\rangle \frac{Var(N_{S})}{\langle N_{S}\rangle
}.
\end{equation}
Thus in superposition models the measured scaled variance of the
multiplicity distribution is a sum of two components. The first
one describes the multiplicity fluctuations from a single source
while the second one accounts for the fluctuations of the number
of sources. In most considerations, fluctuations of the geometry
of the collision process, which in superposition models are given
by the second term in Eq.~\ref{var}, are uninteresting. The
relevant fluctuations are fluctuations determined by the physics
of the collision process, in superposition models given by the
first component of Eq.~\ref{var}. Therefore, in this paper we try
to remove the influence of the second component by the procedure
described in Sec.~\ref{s:data}.

%%%%%%%%%%%%%%%%%%%

\begin{figure}[h]
\begin{center}
\includegraphics[width=14cm]{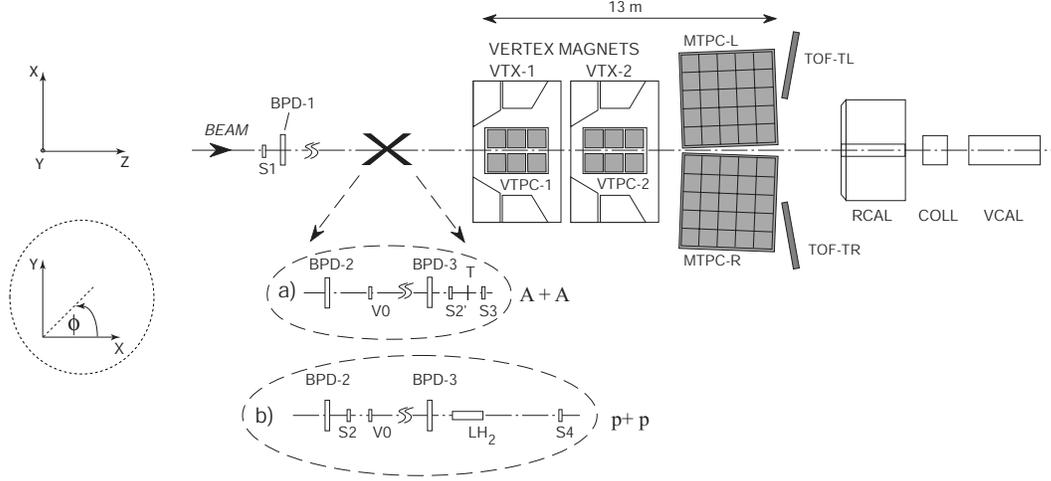}
\end{center}
\caption{\label{setup}The experimental set-up of the NA49
experiment with different beam definitions and target
arrangements.}
\end{figure}

\section{NA49 experimental set-up}
\label{s:experiment}

The NA49 experiment is a large acceptance hadron spectrometer at
the CERN-SPS used to study the hadronic final states produced in
collisions of beam particles (p, Pb from the SPS directly and C,
Si from the fragmentation of the primary Pb beam) with a variety
of fixed targets. The main tracking devices are four large volume
Time Projection Chambers (TPCs) (Fig.~\ref{setup}) which are
capable of detecting 80\% of some 1500 charged particles created
in a central Pb+Pb collision at $158\,A$~GeV. Two of them, the
Vertex TPCs (VTPC-1 and VTPC-2), are located in the magnetic field
of two super-conducting dipole magnets ($1.5$ and $1.1$ T,
respectively) and two others (MTPC-L and MTPC-R) are positioned
downstream of the magnets symmetrically to the beam line. The
results presented here are analysed with a global tracking scheme
\cite{na49_global}, which combines track segments that belong to
the same physical particle but were detected in different TPCs.
The NA49 TPCs allow precise measurements of particle momenta $p$
with a resolution of $\sigma(p)/p^2 \cong (0.3-7)\cdot10^{-4}$
(GeV/c)$^{-1}$. The set--up is supplemented by two Time of Flight
(TOF) detector arrays and a set of calorimeters.

\indent The targets: C (561 mg/cm$^{2}$), Si (1170 mg/cm$^{2}$)
discs and Pb (224 mg/cm$^{2}$) foils for ion collisions and a
liquid hydrogen cylinder (length 20 cm) for elementary
interactions are positioned about 80 cm upstream from VTPC-1.

\indent Pb beam particles are identified by means of their charge
as seen by a Helium Gas-Cherenkov counter (S2') and p beam
particles by a 2 mm scintillator (S2). Both of them are situated
in front of the target. The study of C+C and Si+Si reactions is
possible through the generation of a secondary fragmentation beam
which is produced by a primary target (1 cm carbon) in the
extracted Pb-beam. With the beam line momentum set to $316$~GeV/c
a large fraction of all $Z/A = 1/2$ fragments are transported to
the NA49 experiment. On-line selection based on a pulse height
measurement in a scintillator beam counter (S2) is used to select
particles with $Z=6$ (Carbon) and $Z=13,14,15$ (Al, Si, P).
Off-line clean-up is achieved by using in addition the energy loss
measured by beam position detectors (BPD-1/2/3 in
Fig.~\ref{setup}). These detectors consist of pairs of
proportional chambers and are placed along the beam line for a
precise measurement of the transverse positions of the incoming Pb
nuclei.

\indent For p, C and Si beams interactions in the target are
selected by anti-coincidence of the incoming beam particle with a
small scintillation counter (S4) placed on the beam line between
the two vertex magnets. For p+p interactions at $158\,A$~GeV this
counter selects a (trigger) cross section of 28.5 mb out of 31.6
mb of the total inelastic cross section. For Pb-ion beams, an
interaction trigger is provided by anti-coincidence with a Helium
Gas-Cherenkov counter (S3) directly behind the target. The S3
counter is used to select minimum bias collisions by requiring a
reduction of the Cherenkov signal by a factor of about 6. Since
the Cherenkov signal is proportional to $Z^2$, this requirement
ensures that the Pb projectile has interacted with a minimal
constraint on the type of interaction. This setup limits the
triggers on non-target interactions to rare beam-gas collisions,
the fraction of which proved to be small after cuts, even in the
case of peripheral Pb+Pb collisions.

\indent Centrality of the collisions is selected by using
information from a Veto Calorimeter (VCAL), which measures the
energy of the projectile spectator nucleons. The geometrical
acceptance of the Veto Calorimeter is adjusted in order to cover
the projectile spectator region by a proper setting of the
collimator (COLL).

Details of the NA49 detector set-up and performance of tracking
software are described in~\cite{na49_nim}.

\begin{figure}[h]
\begin{center}
\includegraphics[width=8cm]{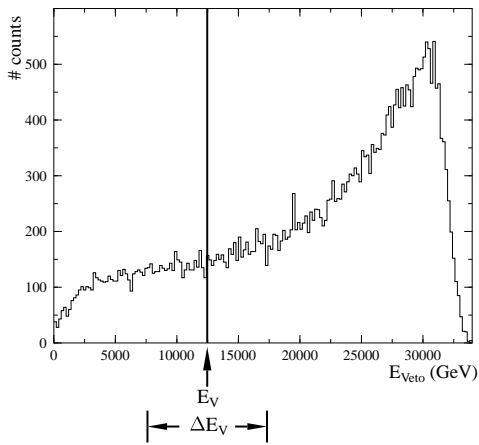}
\end{center}
\caption{\label{veto}Distribution of energy deposited in the Veto
Calorimeter for minimum bias Pb+Pb collisions at $158\,A$~GeV. An
example of a $E_{Veto}$ interval is shown; the interval is
determined by its central value $E_{V}$ and the width $\Delta
E_{V}$.}
\end{figure}

%%%%%%%%%%%%%%%%%%%%%

\section{Data selection and analysis}
\label{s:data}

\subsection{Data sets}
The multiplicity fluctuations are studied for negatively,
positively and all charged particles selecting events within
narrow intervals of energy measured by the Veto Calorimeter
(predominantly energy of projectile spectators). Experimental
material used for the analysis consists of samples of p+p, C+C,
Si+Si and Pb+Pb collisions at $158\,A$~GeV. The number of events
in each sample is given in Table~\ref{no_ev}. For Pb+Pb
interactions a minimum bias trigger was used allowing a study of
centrality dependence.

\begin{table}[h]
\caption{\label{no_ev}The number of events and the fraction of the
total inelastic cross section selected by the on-line trigger for
data sets used in this analysis.}
\begin{center}
\begin{tabular}{lll}
\br
Data Set & No of events & $\sigma/\sigma^{inel}$ \\
\mr
p+p & 319 000 & 0.9\\
C+C & 51 000 & 0.153\\
Si+Si & 59 000 & 0.122\\
Pb+Pb & 65 000 & 0.84\\
\br
\end{tabular}
\end{center}
\end{table}

%%%%%%%%%%%%%%%%%%%%%%%%

\subsection{NA49 acceptance}

The NA49 detector was designed for a large acceptance in the
forward hemisphere. However, also in this region the geometrical
acceptance is not complete. The acceptance limits were
parameterized by a simple function:
\begin{equation}
p_{T}(\phi)=\frac{1}{A+\frac{\phi^{2}}{C}}+B,
\label{eg_acc}
\end{equation}
where the values of the $A$, $B$ and $C$ parameters depend on the
rapidity interval and are given in~\cite{KP}. Only particles
within the analytical curves are used in this analysis. This well
defined acceptance is essential for later comparison of the
results with models and other experiments. Only forward rapidity
tracks ($4.0<y_{\pi}<5.5$, rapidity calculated assuming pion mass
for all particles) with $0.005<p_{T}<1.5$~GeV/c have been used in
this analysis.

%%%%%%%%%%%%%%%%%%%%%%%%%%%%%%

\subsection{Event and particle selection}

The aim of the event selection criteria is to reduce a possible
contamination of non-target collisions. The primary vertex was
reconstructed by fitting the intersection point of the measured
particle trajectories. Only events with a proper quality and
position of the reconstructed vertex were accepted for future
analysis. The vertex coordinate $z$ along the beam had to satisfy
$|z-z_{0}|<\Delta z$, where the nominal vertex position $z_{0}$
and cut parameter $\Delta z$ values are: -579.5 and 5.5 cm, -579.5
and 1.5 cm, -579.5 and 0.8 cm, -578.9 and 0.4 cm for p+p, C+C,
Si+Si and Pb+Pb collisions, respectively. The vertex position in
the transverse $x$, $y$ coordinates had to agree with the incoming
beam position as measured by the BPD detectors.

In order to reduce the contamination of particles from secondary
interactions, weak decays and other sources of non-vertex tracks,
several track cuts were applied. The accepted particles were
required to have measured points in at least one of the Vertex
TPCs. A cut on the extrapolated distance of closest approach of
the particle at the vertex plane has been applied ($|d_{x}|<4$~cm
and $|d_{y}|<2$~cm) to reduce the contribution of non-vertex
particles. Moreover the particle was accepted only when the
potential number of points (calculated on the basis of the
geometry of the track) in the detector exceeded 30. The ratio of
the number of points on a track to the potential number of points
had to be higher than 0.5 in order to avoid split tracks (double
counting).

%%%%%%%%%%%%%%%%%%%%%%%%%

\subsection{Centrality selection}

In order to reduce the effect of fluctuations of the number of
participants the multiplicity fluctuations were analysed for
narrow centrality bins defined by the energy measured in the Veto
Calorimeter.

For the C+C and Si+Si interactions one narrow centrality bin was
selected. In case of Pb+Pb collisions, eight narrow centrality
bins were chosen. The first ($Pb(1)$) bin corresponds to the most
central Pb+Pb events, the last ($Pb(8)$) to the most peripheral
reactions. For each bin of centrality the number of projectile
participants $N_{P}^{PROJ}$ was estimated by:
\begin{equation}
N_{P}^{PROJ}=A-\frac{E_{Veto}}{E_{LAB}}
\label{eqnpproj}
\end{equation}
where $E_{Veto}$ is the energy deposited in the Veto Calorimeter;
$E_{LAB}$ is the energy carried by single nucleon.

The Table~\ref{npproj} shows the number of projectile participants
in each centrality bin.

\begin{table}[h]
\caption{\label{npproj}Data sets used for analysis. Listed for
p+p, C+C, Si+Si and eight centrality bins of Pb+Pb collisions at
$158\,A$~GeV are: the number of projectile participants
$N_{P}^{PROJ}$, the center of each $E_{Veto}$ interval, $E_{V}$,
and the number of events $N_{ev}$ in the interval of width $\Delta
E_{V}=100$~GeV.}
\begin{center}
\begin{tabular}{llll}
\br
Data Set  & $N_{P}^{PROJ}$ & $E_{V}$ (TeV) & $N_{ev}$\\
\mr
p+p & 1 & - & 319 014\\
C+C & 8.8 & 0.5 & 1768\\
Si+Si & 21.7 & 1.0 & 665\\
Pb+Pb(8) & 18.1 & 30.0 & 265\\
Pb+Pb(7) & 37.1 & 27.0 & 290\\
Pb+Pb(6) & 43.4 & 26.0 & 341\\
Pb+Pb(5) & 49.8 & 25.0 & 299\\
Pb+Pb(4) & 81.4 & 20.0 & 266\\
Pb+Pb(3) & 113.1 & 15.0 & 202\\
Pb+Pb(2) & 144.7 & 10.0 & 153\\
Pb+Pb(1) & 176.4 & 5.0 & 150\\
\br
\end{tabular}
\end{center}
\end{table}

\subsection{Multiplicity distributions}

The multiplicity distribution depends on the selected $E_{Veto}$
interval (its position - $E_{V}$ and width - $\Delta E_{V}$; see
Fig.~\ref{veto} for definitions) and the kinematic acceptance
selected for the analysis.

\begin{figure}[h]
\begin{center}
\includegraphics[width=15cm]{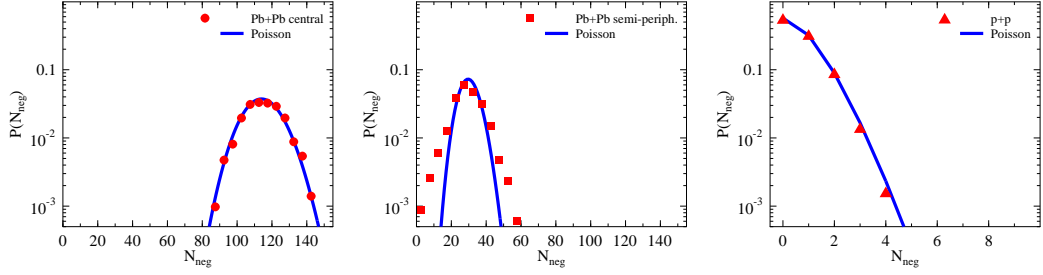}
\end{center}
\caption{\label{mult3}Multiplicity distributions of negatively
charged particles produced in collisions at $158\,A$~GeV, obtained
for $\Delta E_{V}=100$~GeV for central Pb+Pb  collisions with
number of projectile participants $N_{P}^{PROJ}=176$ (left panel);
semi-peripheral Pb+Pb collisions with $N_{P}^{PROJ}=43$
(mid-panel) and p+p interactions with $N_{P}^{PROJ}=1$ (right
panel).}
\end{figure}

In the centrality intervals and acceptance selected for this
analysis, multiplicity distributions show Poissonian behavior for
p+p and central Pb+Pb collisions (see  Fig.~\ref{mult3}). For
semi-peripheral collisions the multiplicity distribution is
significantly broader than the Poisson one.

\begin{figure}[h]
\begin{minipage}{18pc}
\includegraphics[width=18pc]{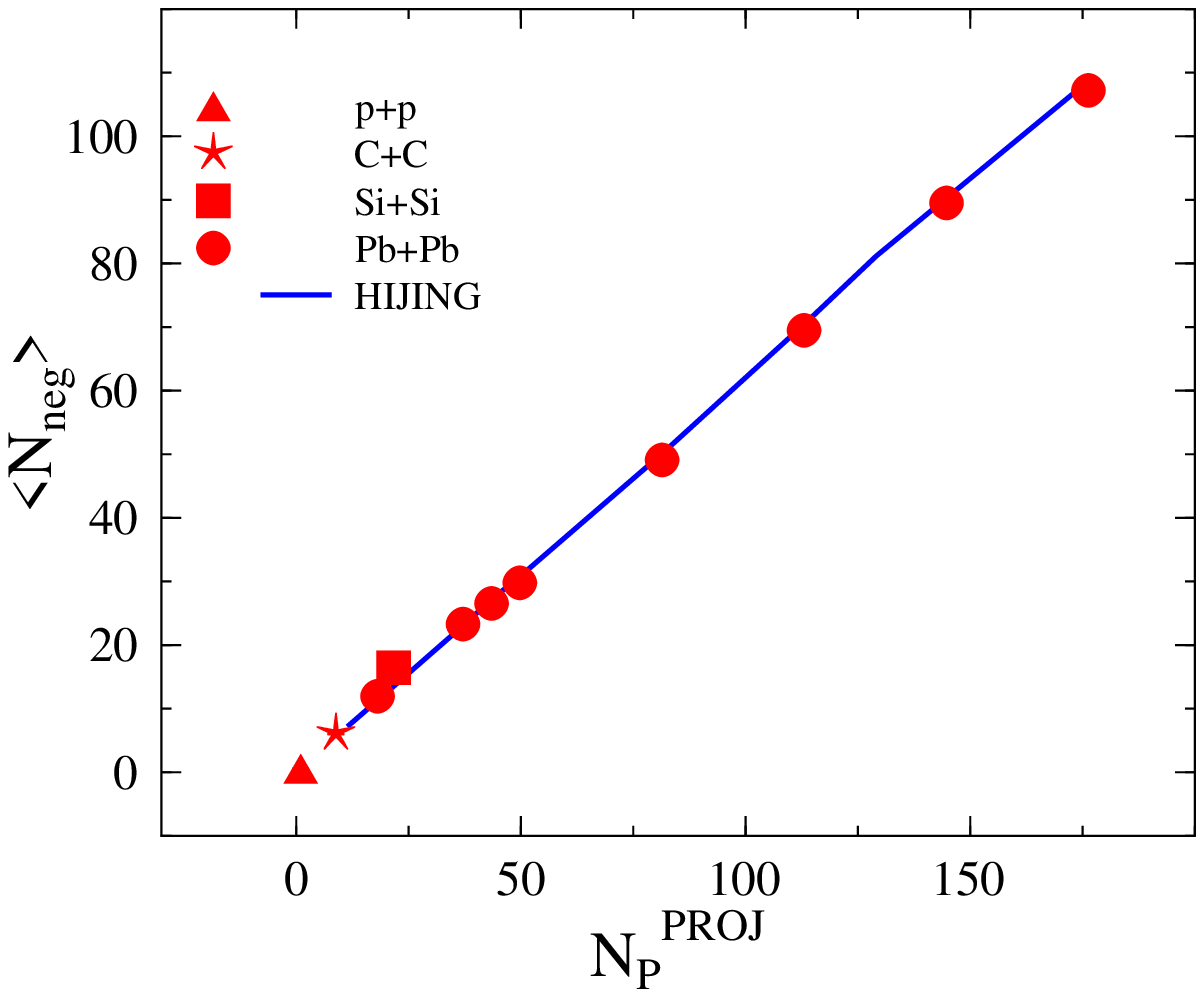}
\caption{\label{mean_mult}The measured mean value of the
multiplicity distribution for negatively charged particles as a
function of the number of projectile participants in comparison
with HIJING simulation in the NA49 acceptance.}
\end{minipage}\hspace{2pc}%
\begin{minipage}{18pc}
\includegraphics[width=18pc]{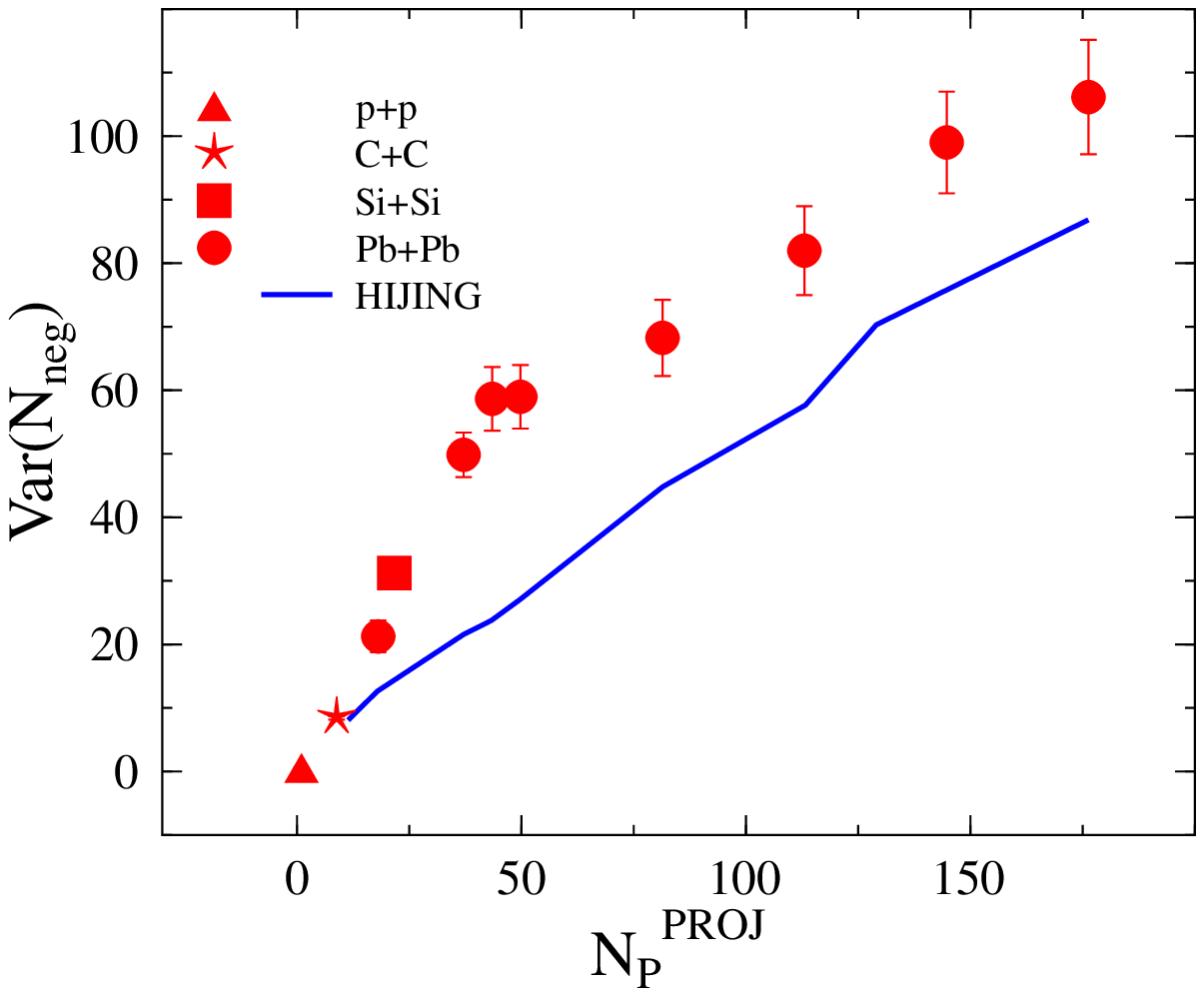}
\caption{\label{var_mult}The measured variance of the multiplicity
distribution for negatively charged particles produced in
collisions at $158\,A$~GeV as a function of the number of
projectile participants in comparison with HIJING simulation in
the NA49 acceptance.}
\end{minipage}
\end{figure}

In Figs.~\ref{mean_mult} and \ref{var_mult} the measured mean
value and the variance of the multiplicity distributions as a
function of the number of projectile participants $N_{P}^{PROJ}$
are presented. The data are compared with the results from the
HIJING model. The simulation was performed with special care taken
to properly grantee the Veto Calorimeter response and the NA49
acceptance. The $E_{Veto}$ energy of a HIJING event was calculated
as the energy of the projectile spectators smeared by a Gaussian
distribution with width
$\sigma\left(E_{Veto}\right)=2\sqrt{E_{Veto}}$ ($E_{Veto}$ is
given in GeV), to take into account detector resolution. One can
see from these plots that the mean multiplicity shows
approximately linear dependence on the number of projectile
participants in data as well as in simulation, whereas the
variance of the multiplicity distributions calculated from the
data exceeds the variance obtained from HIJING (which is roughly a
superposition model).

%%%%%%%%%%%%%%%%%%%%%

\subsection{Scaled variance of the multiplicity distribution}

\begin{figure}[h]
\begin{center}
\includegraphics[width=7cm]{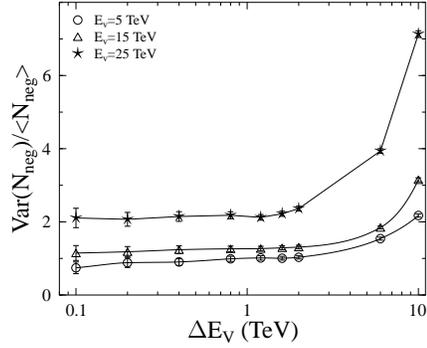}
\end{center}
\caption{\label{mult_delta}Measured scaled variance of
multiplicity distributions of negatively charged particles for
Pb+Pb minimum bias collisions at $158\,A$~GeV as a function of the
interval width $\Delta E_{V}$ of the selected Veto Calorimeter
energy for various positions $E_{V}$ of the interval.}
\end{figure}

The scaled variance of the multiplicity distribution depends on
the width of the energy interval $\Delta E_{V}$ selected in the
Veto Calorimeter. For very broad $\Delta E_{V}$ intervals the
measured scaled variance of the multiplicity distribution has a
large value because of significant fluctuations in number of
projectile participants $N_{P}^{PROJ}$. Narrowing of the $\Delta
E_{V}$ interval results in decreasing fluctuations in the number
of projectile participants and consequently the scaled variance
decreases, as shown in Fig.~\ref{mult_delta}. For $\Delta E_{V}$
smaller than about $1$~TeV the measured scaled variance of the
multiplicity distribution is independent of $\Delta E_{V}$. Note,
that even for very small values of $\Delta E_{V}$ the number of
projectile spectators fluctuates due to the finite resolution of
the Veto Calorimeter.

\indent The scaled variance calculated in the $E_{Veto}$ intervals
defined in Table~\ref{npproj} was corrected for fluctuations in
the number of projectile participants due to the finite width of
the $E_{Veto}$ interval and the finite resolution of the Veto
Calorimeter. Within superposition models the correction $\delta$
can be calculated as:
\begin{equation}
\delta=\frac{\langle N\rangle \cdot
\bigl(Var_{\Delta}\left(E_{Veto}\right)+Var_{R}\left(E_{Veto}\right)\bigr)}{\bigl(E_{BEAM}-\langle
E_{Veto}\rangle \bigr)^{2}},
\end{equation}
where $Var_{\Delta}\left(E_{Veto}\right)$ is the variance of
$E_{Veto}$ due to the finite width of the $E_{Veto}$ bin,
$Var_{R}\left(E_{Veto}\right)$ is the variance of $E_{Veto}$ due
to the finite Veto Calorimeter resolution, $\langle
E_{Veto}\rangle$ is the mean value of $E_{Veto}$ in the bin and
$E_{BEAM}=158\,A$~GeV is the total beam energy.

\begin{figure}[h]
\begin{center}
\includegraphics[width=8cm]{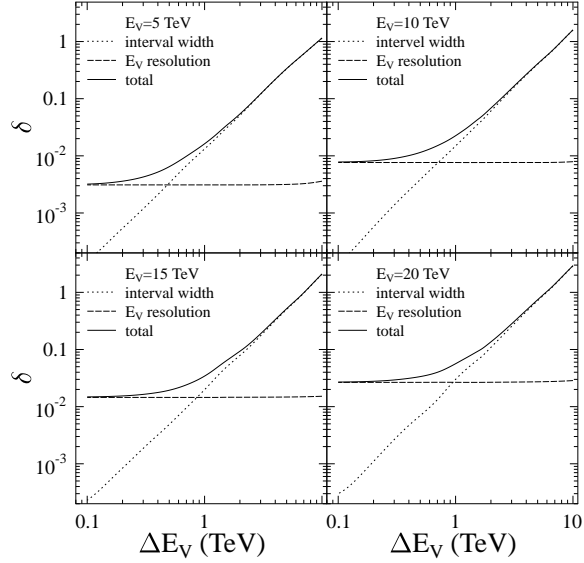}
\end{center}
\caption{\label{corr}The corrections for Veto Calorimeter
resolution and finite interval width as a function of interval
width $\Delta E_{V}$ for various positions $E_{V}$ of the
interval.}
\end{figure}

\begin{figure}[h]
\begin{center}
\includegraphics[width=8cm]{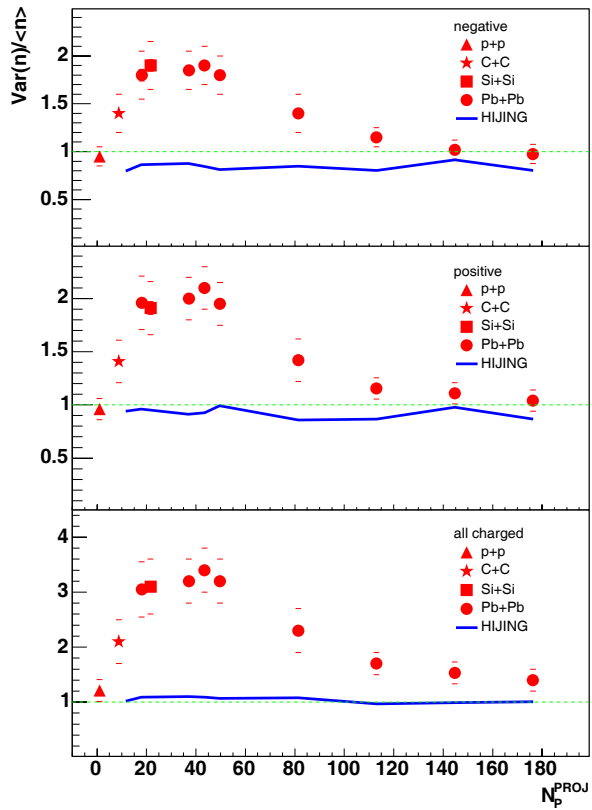}
\end{center}
\caption{\label{varn_all}The scaled variance of the multiplicity
distribution for negatively (upper panel), positively (middle
panel) and all (bottom panel) charged particles produced in
nuclear collisions at $158\,A$~GeV as a function of the number of
projectile participants in comparison with HIJING simulation in
the NA49 acceptance. The error bars represent the systematic
uncertainties, the statistical errors are smaller than the
symbols.}
\end{figure}

\noindent Finally the corrected scaled variance is calculated as:
\begin{equation}
\frac{Var\left(n\right)}{\langle
n\rangle}=\frac{Var\left(N\right)}{\langle N\rangle}-\delta,
\label{eq_main}
\end{equation}
where $Var\left(N\right)$ is the measured variance and $\langle
N\rangle$ the measured mean value of the multiplicity distribution
in a given $E_{Veto}$ bin and $\delta$ represents the corrections
for fluctuations in the number of projectile participants and
finite Veto Calorimeter resolution. The corrections for the data
samples defined in Table~\ref{npproj} are small as shown in
Fig.~\ref{corr}. The largest correction is about $5\,\%$ for the
most peripheral Pb+Pb collisions. The corrected values of the
scaled variance from Eq.~\ref{eq_main} are plotted in
Fig.~\ref{varn_all}.

The systematic error due to Veto Calorimeter resolution,
contamination of non-vertex interactions, tracks from weak decays
and secondary interactions as well as reconstruction
inefficiencies and biases were estimated by varying event and
track selection cuts and simulations. The resulting estimates are
shown by the horizontal bars in Fig.~\ref{varn_all}.

%%%%%%%%%%%%%%%%%%%%%%%%%%%%%

\section{Preliminary results}
\label{s:results}

The results discussed in this section refer to accepted particles,
i.e. particles that are registered by the detector and pass all
kinematic cuts and track selection criteria. The data cover a
broad range in $p_{T}$, ($0.005<p_{T}<1.5$~GeV/c). The rapidity of
accepted particles is restricted to the interval $4.0<y_{\pi}<5.5$
which corresponds to forward rapidities in the collision of equal
mass nuclei (at $158\,A$~GeV energy the center of mass rapidity
equals $2.9$ for fixed target geometry), where the azimuthal
acceptance is large. The acceptance in azimuthal angle is given by
Eq.~\ref{eg_acc}.

The corrected scaled variance of the multiplicity distribution for
negatively, positively and all charged accepted particles as a
function of centrality in comparison with HIJING simulation is
shown in Fig.~\ref{varn_all} (the results were also shown at the
Quark Matter 2004 conference \cite{Gazdzicki}). HIJING produces
approximately Poissonian multiplicity distributions, independent
of centrality. In contrast, the data points indicate a
non-monotonic dependence on system size with a maximum at number
of projectile participants $N_{P}^{PROJ}\simeq 35$. Note, that the
value of unity for p+p interactions is an accident at
$158\,A$~GeV. The multiplicity distribution in p+p collisions is
not Poissonian at lower and higher energies; see for
example~\cite{Gazdzicki:1990bp}.

Within a statistical model in which strict electric charge
conservation is obeyed \cite{Begun} the scaled variance of
like-sign particles is expected to vary in the range 0.5-1
depending on the volume of the matter and the acceptance in
momentum space. The results presented here are in disagreement
with this prediction.

The scaled variances for positively and negatively charged
particles are similar. The corresponding values for all charged
particles are larger. This is probably due to charge conservation.
Assuming that negatively and positively charged particles in the
experimental acceptance are correlated with a correlation factor
$\rho$, one gets:
\begin{equation}
\frac{Var\left(n_{ch}\right)}{\langle n_{ch}\rangle}=
\frac{Var\left(n_{neg}\right)}{\langle
n_{neg}\rangle}\left(1+\rho\right)=\frac{Var\left(n_{pos}\right)}{\langle
n_{pos}\rangle}\left(1+\rho\right)
\end{equation}
where $n_{neg}$, $n_{pos}$ and $n_{ch}$ are multiplicities of
negatively, positively and all charged particles, respectively,
and, for simplicity we assumed $\langle n_{neg}\rangle=\langle
n_{pos}\rangle$.

\noindent The scaled variance for charged particles is $1+\rho$
times larger than for like-sign particles (see also \cite{Begun}).

Note that transverse momentum fluctuations measured in nuclear
collisions at $158\,A$ GeV \cite{KP} also show a non-monotonic
system size dependence with the maximum located close to the
maximum of the scaled variance of multiplicity distribution. A
possible relation between these two observation was discussed in
\cite{Mrowczynski:2004cg}.

%%%%%%%%%%%%%%%%%%%%%%%%%

\section{Summary}
\label{s:summary}

The scaled variance of the multiplicity distribution was used as a
measure of multiplicity fluctuations. A non-monotonic system size
dependence of the scaled variance is seen for negatively,
positively and all charged particles. The scaled variance is
closest to unity for p+p and central Pb+Pb collisions and has a
maximum at $N_{P}^{PROJ}\simeq 35$. The behavior of the scaled
variance is similar for positively and negatively charged
particles, but it is larger for all charged particles.

%%%%%%%%%%%%%%%%%%%%%%

\section*{Acknowledgements}

This work was supported by the US Department of Energy Grant
DE-FG03-97ER41020/A000, the Bundesministerium fur Bildung und
Forschung, Germany, the Polish State Committee for Scientific
Research (2 P03B 130 23, SPB/CERN/P-03/Dz 446/2002-2004, 2 P03B
04123), the Hungarian Scientific Research Foundation (T032648,
T032293, T043514), the Hungarian National Science Foundation,
OTKA, (F034707), the Polish-German Foundation, and the Korea
Research Foundation Grant (KRF-2003-070-C00015).

\end{document}